\documentclass{article}
\usepackage[T1]{fontenc}
\usepackage[utf8]{inputenc}
\usepackage[]{ismir} 
\usepackage{amsmath,cite,url}
\usepackage{graphicx}
\usepackage{color}

\usepackage{amsfonts}
\usepackage{amssymb}
\usepackage{graphicx}
\usepackage{multirow}
\usepackage{booktabs}
\usepackage{ctable}
\usepackage{csquotes}

\title{The Rhythm In Anything: Audio-Prompted Drums Generation with Masked Language Modeling}

\multauthor
 {Patrick O'Reilly$^1$ \hspace{1cm} Julia Barnett$^1$ \hspace{1cm} Hugo Flores Garcia$^1$}
{{\bf Annie Chu$^1$ \hspace{1cm} Nathan Pruyne$^1$ \hspace{1cm} Prem Seetharaman$^2$ \hspace{1cm} Bryan Pardo$^1$}\\
$^1$Northwestern University, Evanston, USA \hspace{1cm} $^2$Adobe Research, San Francisco, USA\\
{\tt\small patrick.oreilly2024@u.northwestern.edu}
}

\def\authorname{P. O'Reilly, J. Barnett, H. F. Garcia, A. Chu, N. Pruyne, P. Seetharaman, and B. Pardo}

\usepackage[bookmarks=false,pdfauthor={\authorname},pdfsubject={\pdfsubject},hidelinks]{hyperref}

\begin{document}

\maketitle

\begin{abstract}
Musicians and nonmusicians alike use rhythmic sound gestures, such as tapping and beatboxing, to express drum patterns. 
While these gestures effectively communicate musical ideas, realizing these ideas as fully-produced drum recordings can be time-consuming, potentially disrupting many creative workflows. To bridge this gap, we present TRIA (\textbf{T}he \textbf{R}hythm \textbf{I}n \textbf{A}nything), a masked transformer model for mapping rhythmic sound gestures to high-fidelity drum recordings. Given an audio prompt of the desired rhythmic pattern and a second prompt to represent drumkit timbre, TRIA produces audio of a drumkit playing the desired rhythm (with appropriate elaborations) in the desired timbre. Subjective and objective evaluations show that a TRIA model trained on less than 10 hours of publicly-available drum data can generate high-quality, faithful realizations of sound gestures across a wide range of timbres in a zero-shot manner.

\end{abstract}

\section{Introduction}\label{sec:introduction}

\begin{figure}
  \centering
  \includegraphics[alt={TRIA hero diagram},width=0.9\linewidth]{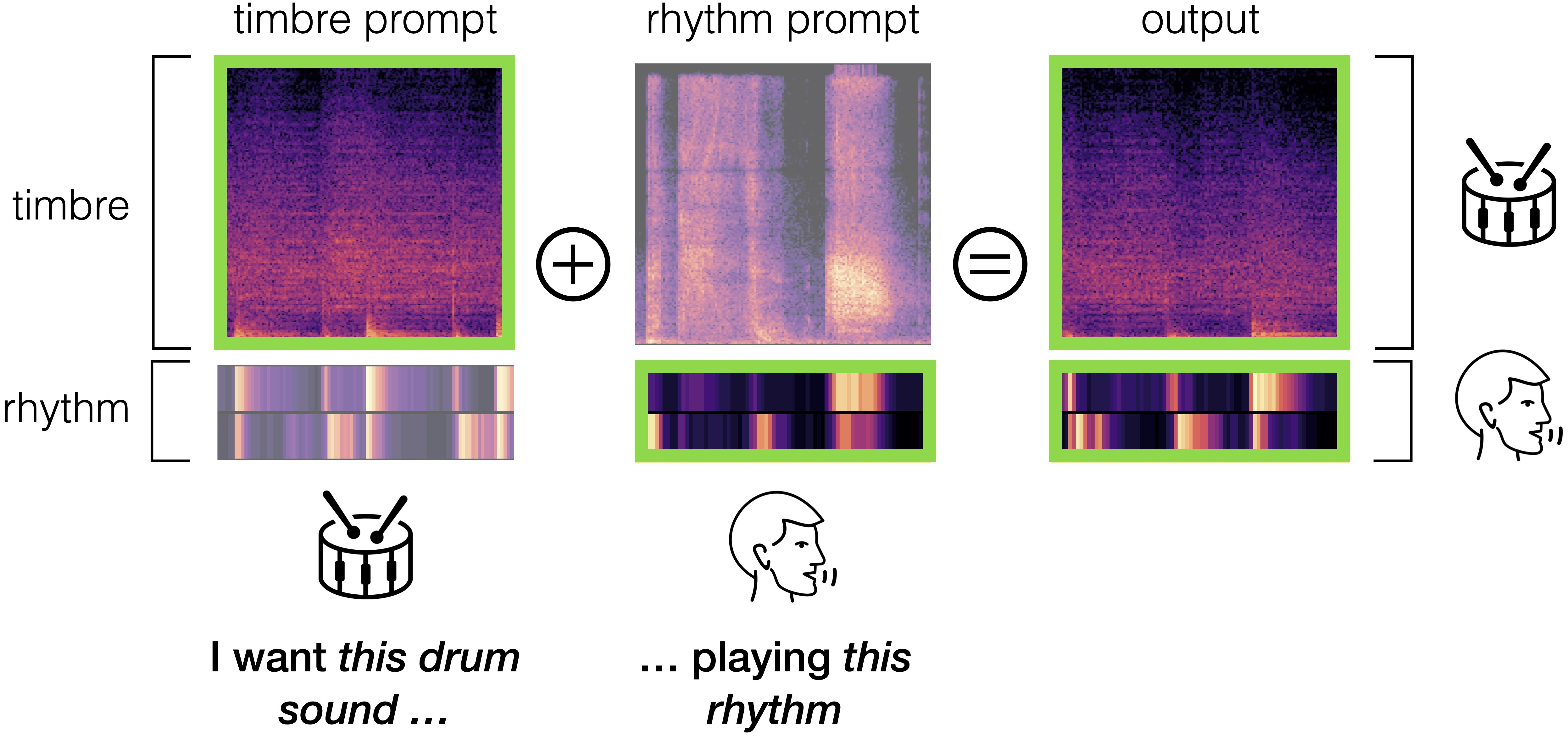}
  \caption{TRIA conditions generation of a new drum recording on two prompts: the timbre of an example drum recording (illustrated by a spectrogram), and the rhythm of a sound gesture (the dualized features in Section \ref{subsec:feats}).}
  \label{fig:hero}
\end{figure}

Sound gestures such as tapping and beatboxing provide a convenient and idiomatic means of expressing rhythmic ideas. 
Rather than ``literally'' specifying a rhythmic idea through one-to-one imitation, sound gestures often capture a reduced, high-level representation of the desired rhythm---for instance, a beatboxer may only voice one element where many have simultaneous onsets, or leave certain elements unvoiced and implied.
Realizing these gestures as fully-produced drum arrangements often requires many steps: the voiced sound elements in a gesture must be mapped to appropriate drum parts, unvoiced or implied elements must be plausibly reconstructed, the resulting arrangement must be performed and recorded or sequenced and synthesized digitally in audio editing software, and the final recording may require further processing to shape the timbre satisfactorily. By contrast, many creative workflows may benefit from the ability to rapidly generate diverse full-drumkit realizations of rhythmic sound gestures.

To bridge this gap, we propose TRIA (\textbf{T}he \textbf{R}hythm \textbf{I}n \textbf{A}nything), a masked transformer model for mapping arbitrary rhythmic sound gestures to high-fidelity drum recordings. Given two audio prompts---one specifying the basic desired rhythm via a sound gesture, and one specifying the desired drum timbre via an example recording---TRIA synthesizes full-drumkit audio playing a fleshed-out arrangement of the desired rhythm in the desired timbre. 
TRIA can faithfully realize sound gestures in unseen timbres in a zero-shot manner despite its relatively small model size (43M trainable parameters) and training dataset (less than 10 hours of publicly-available drum recordings from MusDB18-HQ \cite{musdb}). Through both quantitative comparisons and qualitative human listening evaluations, we demonstrate that TRIA matches or exceeds the performance of a 1-billion parameter state-of-the-art model \cite{melodyflow} trained on $20,000$ hours of public and private data in converting sound gestures to drum recordings.

\begin{figure*}
  \centering
  \includegraphics[alt={TRIA system diagram},width=0.9\linewidth]{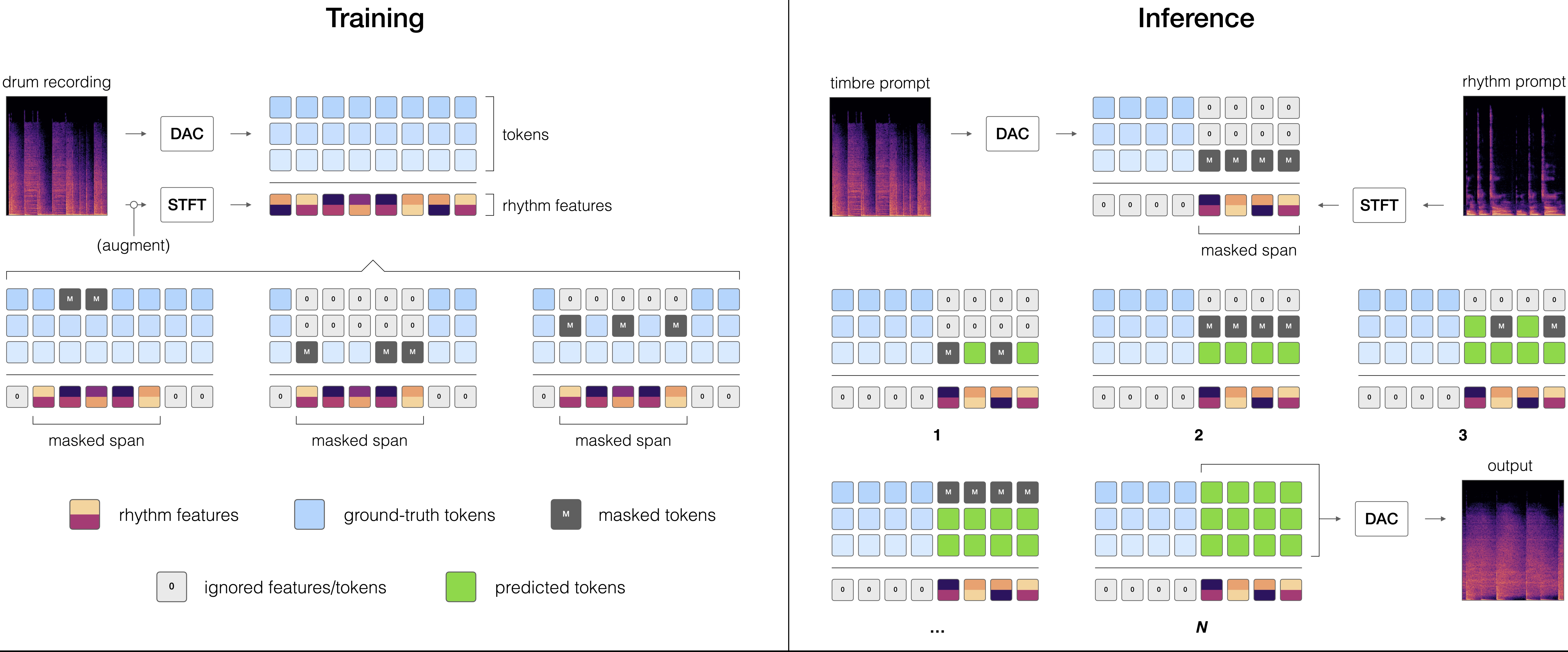}
  \caption{The proposed TRIA system. During training (left), acoustic tokens of a tokenized drum recording are predicted, conditioned on surrounding unmasked tokens and rhythm features extracted from an augmented version of the recording; we illustrate three training examples. During inference (right), we fix the timbre prompt as a prefix and predict a masked suffix conditioned on aligned features extracted from the rhythm prompt. Inference predicts tokens in coarse-to-fine order.
  }
  \label{fig:main}
\end{figure*}

Our contributions are as follows:
\begin{enumerate}
    \item A model capable of mapping arbitrary rhythmic sound gestures to high-fidelity drum recordings using drum timbres specified at inference time
    \item A dualized representation that lets the model capture salient rhythmic structure across drum and non-drum sound classes
    \item Subjective and objective evaluations showing the importance of the dualized representation and the model's ability to generate musically-pleasing translations that adhere to rhythm and timbre prompts

\end{enumerate}

\noindent We provide audio examples and code on our webpage.\footnote{\url{https://therhythminanything.github.io/}}

\section{Related Work}

The translation of simple rhythmic gestures into full drum beats has been explored in the symbolic domain, notably in the GrooVAE models proposed by Gillick et al. \cite{groovae}. While these allow for mapping single-voice MIDI drum patterns to full-drumkit expressive MIDI performances, they do not allow for audio-prompted rhythm or timbre specification. 

In the audio domain, Santos \& Cardoso applied RAVE \cite{rave} models to a tap-to-drums translation task \cite{fromtapstodrums}. However, RAVE does not support zero-shot audio-prompted timbre specification, but instead requires re-training for each new specified timbre.
In general, recent neural network-based timbre transfer systems are similarly constrained or else support only pitched instruments \cite{ddsp, after}. One exception is MelodyFlow \cite{melodyflow}, which performs text-guided audio editing via latent diffusion inversion, hypothetically allowing for the specification of arbitrary timbres via text prompts. We perform extensive comparisons between our proposed system and MelodyFlow in Section \ref{sec:experiments}.

A number of systems use transcription to translate beatbox audio into drum recordings via synthesis from a predicted MIDI representation, but generally require user-specific calibration for accurate transcription and do not support audio-prompted timbre specification \cite{lvt, dubler2}. In general, transcription-based systems are constrained to narrow sound gesture types with well-defined audio-symbol mappings or available annotated data for supervised training (e.g., beatboxing), and limited to ``literal'' mappings of timbres onto atomic sound events. By contrast, we propose an audio-prompted, \textit{self-supervised} approach for mapping simple rhythmic gestures to potentially complex full-drumkit recordings, allowing for the generation of arrangement details not explicit in the rhythm gesture.

Previous works have hypothesized that musicians often perceive and arrange drum patterns using implicit two-voice ``dualized'' representations that oscillate between low and high states \cite{bistatereduction, taptamdrum}. However, the use of dualized representations for music generation has been limited to the symbolic domain \cite{dualizationofrhythmpatterns}. Our proposed system obtains dualized representations from audio (Section \ref{subsec:feats}) to guide the generation of drum audio, letting us specify rhythmic structure with non-drum sounds (e.g. finger tapping).

Finally, our work differs from prior work on generating symbolic rhythm patterns \cite{selfsim, eigenrhythms, networkrepresentations}, drum loops \cite{deepdrummer, drumnet}, and drum samples \cite{drumgan, sylewavegan, ganinversion} in that we seek to convert sound gestures into audio-domain full drumkit performances.

\section{Method}\label{sec:method}

We next describe the design of the proposed TRIA system.

\textbf{Architecture: } Similar to VampNet \cite{vampnet}, TRIA is a transformer-based masked language model. TRIA consists of 12 standard transformer blocks, each with hidden size $h = 512$, $8$ attention heads, and rotary positional encoding \cite{rope}, resulting in $43$ million trainable parameters.

\textbf{Audio Tokenization: }
TRIA predicts acoustic tokens produced by Descript Audio Codec (DAC) \cite{dac}. Within DAC, audio is segmented into a series of $T$ frames, each of which is mapped to a vector representation via a fully convolutional encoder. Encoded vectors are quantized with a hierarchical sequence of $C$ vector-quantizers, each with its own codebook. Each quantizer encodes the residual between the original and the quantized representation produced by the previous quantizers. Quantized vectors are represented by their codebook indices, resulting in a token representation of $C$ codebooks by $T$ frames. A matched decoder converts $C \times T$ token representations into audio.

\textbf{Masked Language Modeling: }TRIA generates drum audio by predicting missing or ``masked'' DAC tokens within a partially-masked ``buffer'' of size $C \times T$, conditioned on unmasked tokens (representing the target timbre and generated content), as is typical for masked token modeling. TRIA, however, also conditions generation on aligned rhythm features  representing the target rhythm (see Section \ref{subsec:feats}). Once all masked tokens are predicted, they are mapped to 44.1kHz mono audio via the DAC decoder.

To produce predictions for masked tokens in the buffer within a specific codebook $c \in [0, C - 1]$, all tokens in the buffer are first mapped to continuous vectors of size $h$ via separate learned embedding tables per codebook, with masked tokens mapped to a single learned ``mask'' embedding shared across all codebooks. Recall that the tokens in every codebook at level $c' > c$ correct the residual error of the token at level $c$. Therefore, if a token at level $c$ is masked, all corresponding embedding vectors in codebooks $c' > c$ are zeroed. Embedding vectors are then summed across codebooks to obtain a sequence of shape $h \times T$. Rhythm features (Section \ref{subsec:feats}) are projected to the hidden dimension and zeroed for frames in which there are no masked tokens, resulting in a corresponding conditioning sequence of shape $h \times T$. The two sequences are summed and passed to the transformer, which predicts a probability distribution over tokens in codebook $c$ at each frame via one of $C$ codebook-specific projection layers.

\textbf{Inference: } At inference, we take as inputs a timbre prompt (drum) recording and a rhythm prompt (sound gesture) recording. We construct a buffer in which the tokenized timbre prompt serves as an unmasked prefix, with all subsequent frames (corresponding to the length of the rhythm prompt) fully masked. We compute rhythm features aligned to this masked suffix from the rhythm prompt.  

We then perform SoundStorm-style inference \cite{soundstorm} to iteratively predict masked tokens in each codebook in coarse-to-fine order, using the schedule of Chang et al. \cite{maskgit} to gradually unmask or ``confirm'' tokens in the suffix. We adopt temperature-based nondeterministic unmasking from VampNet and causal bias from StemGen \cite{stemgen} to favor unmasking earlier tokens in the buffer first.

Thus, we fill in the masked suffix using timbral information from the timbre prompt and rhythmic information from the rhythm prompt, resulting in a generation that adheres to both prompts. By specifying the number of inference iterations over which each codebook is unmasked, we can expend more compute on challenging high-entropy early generation steps and less on highly-determined later steps. For all experiments reported in this paper, we use an inference schedule of $\{8, 8, 8, 8, 8, 4, 4, 4, 4 \}$ iterations for DAC's $9$ respective codebooks in coarse-to-fine order, classifier-free guidance \cite{classifierfreeguidance} weight $2.0$, unmasking temperature $10.0$, and causal bias $1.0$.

\textbf{Training: } At each training iteration, we sample a drum recording that serves as both timbre and rhythm prompt, tokenizing with DAC to obtain a buffer and computing rhythm features at a matching temporal resolution. We select a random codebook and a random span of consecutive frames covering 50\% to 75\% of the buffer length, and mask a subset of tokens within this codebook and span according to the cosine schedule proposed by Chang et al. \cite{maskgit}; we then compute cross-entropy loss between TRIA's predicted distributions at masked token positions and the corresponding ground-truth tokens.  
To allow TRIA to process rhythm prompts from a variety of sound sources and recording conditions, we apply noise, band-pass filtering, pitch shift, phase shift, and equalization to the rhythm prompt audio with independent $25$\% probabilities at each iteration. To provide control over the degree of adherence to the rhythm prompt, we implement classifier-free guidance \cite{classifierfreeguidance} by zeroing rhythm features in 20\% of training iterations to learn unconditional mappings, and then performing weighted interpolation between unconditional and conditional predictions at inference time. 

We train all TRIA models on drums from a 90\% split of the MusDBHQ-18 dataset \cite{musdb}, totaling 8 hours of audio. We train on 6-second random excerpts for $100,000$ iterations at a batch size of $48$ on $4 \times$ NVIDIA A10G GPUs, requiring $\sim27$ hours per model. Training and inference are illustrated in Figure \ref{fig:main}.

\begin{figure}
  \centering
  \includegraphics[alt={Subjective evaluation},width=0.9\linewidth]{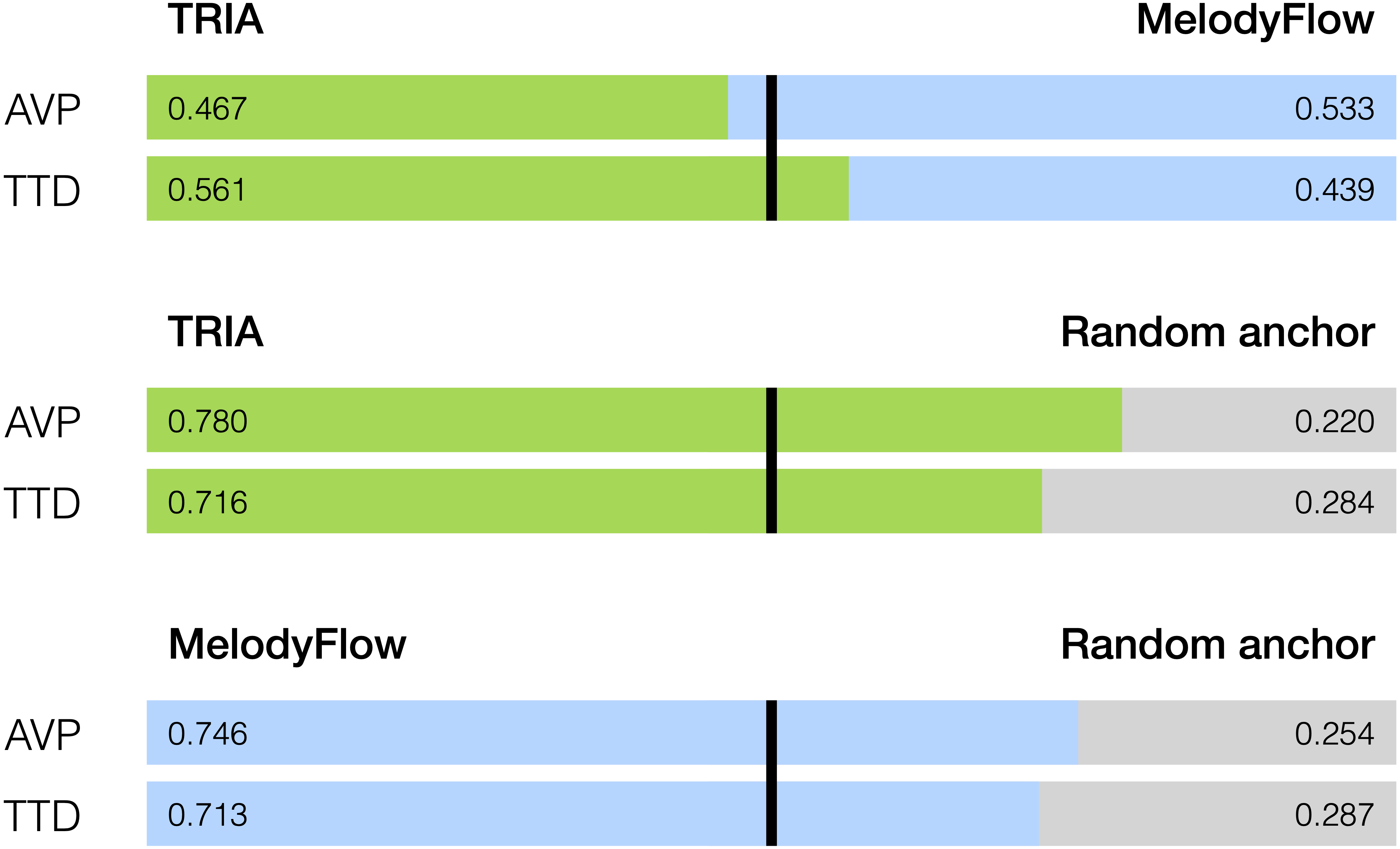}
  \caption{Results of the listener preference evaluation detailed in Section \ref{subsec:survey}. We plot win rates for TRIA and MelodyFlow generations from rhythm prompts sampled from the AVP and TapTamDrum (TTD) datasets, as well as random anchors from MoisesDB drums. 
  }
  \label{fig:survey}
\end{figure}

\begin{table}
\centering
\renewcommand{\arraystretch}{1.2}
\addtolength{\tabcolsep}{-0.08em}
\begin{tabular}{l|cc|cc}
 & \multicolumn{2}{c|}{\textbf{F1 Snare} $\uparrow$} & \multicolumn{2}{c}{\textbf{F1 Kick} $\uparrow$} \\
\textbf{Model} & 30ms  & 100ms & 30ms & 100ms \\
\specialrule{1.5pt}{0pt}{0pt}
Random anchor & 0.04 & 0.15 & 0.09 & 0.29  \\
\hline
MelodyFlow\textsubscript{0.0} & 0.08 & 0.16 & 0.11 & 0.19 \\
MelodyFlow\textsubscript{0.1} & 0.11 & 0.13 & 0.13 & 0.18 \\
MelodyFlow\textsubscript{0.2} & 0.19 & 0.23 & 0.21 & 0.23 \\
\hline
TRIA\textsubscript{1Band} & 0.23 & 0.35 & 0.38 & 0.50 \\
TRIA\textsubscript{2Band}* & 0.32 & 0.47 & 0.52 & 0.66 \\
TRIA\textsubscript{2Band-NA} & 0.10 & 0.17 & 0.47 & 0.62 \\
TRIA\textsubscript{3Band} & 0.33 & 0.50 & 0.61 & 0.71 \\
TRIA\textsubscript{4Band} & 0.30 & 0.47 & 0.59 & 0.72 \\
\end{tabular}
\caption{F1 scores of automatic snare and kick transcriptions of MelodyFlow and TRIA generations from annotated AVP beatbox recordings at $30$ms and $100$ms onset tolerances. Higher scores indicate generations preserve the placement of kicks and snares from beatbox recordings.}
\label{tab:transcription}
\end{table}
\vspace{-1.0em}


\subsection{Dualized Rhythm Representation}\label{subsec:feats}

To allow inference on arbitrary sound gestures while training only on drum audio, we require (1) timbre-rhythm disentanglement, with timbre information for the prediction of masked token spans provided by unmasked tokens outside the span and rhythm information provided by aligned rhythm features within the span; and (2) a rhythm feature representation that captures the structure of both drums and sound gestures with vastly different frequency energy distributions. If timbre-rhythm disentanglement is not enforced, e.g. if timbre information leaks from the rhythm features, TRIA will not apply the specified timbre. If there exists a modality gap between drums and sound gestures within the rhythm feature representation, TRIA will struggle to map sound gestures to plausible drum generations.

The simplest rhythm representation satisfying these criteria is a one-dimensional sequence of loudness estimates, which captures onset information similar to GrooVAE \cite{groovae}. However, researchers have found that onset representations fail to adequately capture relationships between multiple elements within percussion patterns and human sound gestures \cite{bistatereduction, taptamdrum} -- for instance, distinct kick and snare vocalizations within a beatbox recording may be ``flattened'' into indistinguishable loudness spikes, making it difficult for TRIA to faithfully map the beatbox to drums. On the other hand, if the rhythm feature representation is too fine-grained, e.g. a full spectrogram, it will leak timbre information from the rhythm prompt and cause TRIA to ignore the timbre prompt. Additionally, drums and sound gestures will likely manifest distinctly in fine-grained feature representations, causing a train-inference mismatch.

To address these potential pitfalls, we propose a rhythm feature representation based on a two-band spectrogram with an adaptive splitting frequency. We start with an 80-bin mel-spectrogram of the rhythm prompt audio and compute a splitting frequency that equally divides energy into low and high bands, summing all bins within each band. We then standardize each band independently, apply a sigmoid nonlinearity to bound all values to $\left[ 0, 1 \right]$, and quantize all values to 33 steps ($0$, $\frac{1}{32}$, $\frac{2}{32}$, ..., $1$) within this range.

Our motivation for this representation is twofold. First, a two-voice representation allows core elements of drum recordings and sound gestures to manifest distinctly, but lacks sufficient detail to leak timbre information or distinguish between drum recordings and sound gestures. Second, two-voice or ``dualized'' rhythm representations have been explored previously for the analysis and generation of drum patterns in the symbolic domain \cite{bistatereduction, taptamdrum, dualizationofrhythmpatterns}. 
We extend this line of inquiry by evaluating the efficacy of audio-derived dualizations for audio generation.

\begin{table}
\centering
\renewcommand{\arraystretch}{1.2}
\addtolength{\tabcolsep}{-0.08em}
\begin{tabular}{l|ccc}
 & \multicolumn{3}{c}{\textbf{MFCC-Sim}} \\
\textbf{Model} & Rhythm $\downarrow$  & Timbre $\uparrow$ & Random \\
\specialrule{1.5pt}{0pt}{0pt}
MelodyFlow\textsubscript{0.0} & 0.88 & - - & 0.81 \\
MelodyFlow\textsubscript{0.1} & 0.92 & - - & 0.86 \\
MelodyFlow\textsubscript{0.2} & 0.96 & - - & 0.85 \\
\hline
TRIA\textsubscript{1Band} & 0.85 & 0.95 & 0.87 \\
TRIA\textsubscript{2Band}* & 0.85 & 0.96 & 0.87  \\
TRIA\textsubscript{2Band-NA} & 0.83 & 0.93 & 0.85  \\
TRIA\textsubscript{3Band} & 0.86 & 0.95 & 0.87  \\
TRIA\textsubscript{4Band} & 0.84 & 0.96 & 0.86  \\
\end{tabular}
\caption{Timbral similarity between model outputs, input rhythm/timbre prompts, and random drum recordings as measured by time-averaged MFCC cosine similarity. Higher-than-random similarity with the rhythm prompt implies timbre leakage, while higher-than-random similarity with the timbre prompt implies prompt adherence.}
\label{tab:timbre}
\end{table}

\section{Experiments}\label{sec:experiments}

We empirically validate TRIA's ability to map sound gestures to full-drumkit recordings in user-specified timbres across two specific sound gesture types (beatboxing and tapping). We conduct both subjective human evaluations and objective evaluations of generation quality and adherence to rhythm and timbre prompts.

\subsection{Models}
\label{subsec:models}

\textbf{TRIA}: In addition to the TRIA system described in Section \ref{sec:method} (TRIA\textsubscript{2Band}*), we validate our choice of rhythm feature representation by comparing variants of TRIA trained on 1-band (TRIA\textsubscript{1Band}), 2-band with no adaptive frequency split (TRIA\textsubscript{2Band-NA}), 3-band (TRIA\textsubscript{3Band}), and 4-band rhythm features (TRIA\textsubscript{4Band}). 

\textbf{MelodyFlow}: we compare TRIA to MelodyFlow \cite{melodyflow}, a state-of-the-art text-prompted music editing system. MelodyFlow can apply text-specified timbres to sound gestures using regularized latent inversion, which maps an encoded sound gesture to an initial noise estimate and then resynthesizes it conditioned on the text prompt via flow-matching. This is done by a 1-billion parameter transformer model trained on a mix of private and licensed music totalling $20,000$ hours. The degree to which MelodyFlow preserves the structure of the rhythm prompt can be coarsely controlled by specifying the ``target flow step'' for inversion, with $0.0$ corresponding to full noising and $1.0$ corresponding to no noising (where the audio is left unaltered). In our experiments we compare target flow steps of $0.0$, $0.1$, and $0.2$ (MelodyFlow\textsubscript{0.0}, MelodyFlow\textsubscript{0.1},  and MelodyFlow\textsubscript{0.2}, respectively); we find that higher values result in negligible adherence to the specified timbre. We use the default settings of 128 inference steps, ``Euler'' solver, and ReNoise \cite{renoise} regularization strength $0.2$. To allow fair comparisons with TRIA, we downmix MelodyFlow generations from stereo to mono and downsample from 48kHz to 44.1kHz.

\subsection{Datasets}

We evaluate both TRIA and MelodyFlow on rhythm prompts drawn from two datasets of sound gestures: AVP \cite{avp}, containing 56 amateur beatbox improvisations across 28 participants and 2 conditions with human-annotated transcriptions; and TapTamDrum \cite{taptamdrum}, containing 1116 two-tone tapping imitations of drum beats across 4 participants. To avoid overlap with TRIA's training data, we sample audio timbre prompts from the MoisesDB dataset \cite{moises}, which contains drum stems from 240 commercial-quality music tracks. Because MelodyFlow requires timbre specification via text rather than audio prompts, we generate 50 descriptions of acoustic and electronic drum kit timbres using GPT-4.5 \cite{chatgpt} which we manually inspect to ensure quality and diversity. Due to the lack of available drumkit timbre description datasets and our difficulty in obtaining diverse captions from drum audio using existing multimodal models \cite{audioflamingo2}, we settle on these synthetic descriptions as a reasonable approximation of ``plausible" text prompts, and consult with the MelodyFlow authors to ensure descriptions are formatted appropriately for the model. In all experiments, we sample 2-second timbre prompts for TRIA and generate from rhythm prompts trimmed to a maximum duration of 4 seconds.

\begin{table}
\centering
\renewcommand{\arraystretch}{1.2}
\addtolength{\tabcolsep}{-0.08em}
\begin{tabular}{l|c|c} \textbf{Model} &
 \textbf{KAD\textsubscript{PANN}} $\downarrow$ & \textbf{KAD\textsubscript{CLAP}} $\downarrow$ \\
\specialrule{1.5pt}{0pt}{0pt}
TRIA\textsubscript{1Band} & 6.95 & 6.81  \\
TRIA\textsubscript{2Band}* & 4.56 & 5.05 \\
TRIA\textsubscript{2Band-NA} & 6.61 & 10.63 \\
TRIA\textsubscript{3Band} & 4.53 & 5.46 \\
TRIA\textsubscript{4Band} & 4.14 & 4.61  \\
\end{tabular}
\caption{Kernel Audio Distance (KAD) between a set of $500$ generations from each model and a reference distribution of $500$ drum excerpts from MoisesDB; lower scores indicate better audio quality.}
\label{tab:quality}
\end{table}

\subsection{Subjective Evaluation}
\label{subsec:survey}

We first aim to understand how human listeners rate TRIA's translations of sound gestures to drums when compared to the state-of-the-art model MelodyFlow. To this end, we conduct a listening evaluation utilizing ReSEval \cite{morrison2022reproducible}, a framework for subjective evaluation tasks  on crowdworker platforms; we recruit evaluators through the online research platform Prolific\footnote{\url{https://www.prolific.com/}}. We evaluate the TRIA\textsubscript{2Band}* and MelodyFlow\textsubscript{0.2} variants, as we find that these models provide a good balance of adherence to both rhythm and timbre prompts.

\textbf{Data Preparation}: We prepared 80 sets of short ($3$--$4$ second) audio clips. Each set contained (1) a reference sound gesture serving as a rhythm prompt, drawn either from the AVP ``personal'' condition (beatboxing) or TapTamDrum (tapping); (2) a TRIA generation from the rhythm prompt; (3) a MelodyFlow generation from the rhythm prompt; and (4) a random MoisesDB drum excerpt, unrelated to the rhythm prompt, as a low anchor. We generated these 80 sets using 10 rhythm prompts (5 beatboxing, 5 tapping) and 8 timbre prompts per rhythm prompt. TRIA's audio timbre prompts were drawn randomly from MoisesDB drum excerpts, while MelodyFlow's text timbre prompts were drawn from the aforementioned set of 50 generated timbre descriptions. To ensure broadly comparable timbres across generations, we restricted our audio timbre prompts to acoustic drum kit recordings and our text prompts to descriptions of acoustic drum kit timbres.

\textbf{ABX Trials}: 
We leveraged the findings of Cartwright et al.\cite{cartwright2016fast, cartwright2018crowdsourced} and deployed pairwise comparison evaluations using remote crowdworkers. In our study, crowdworkers performed ABX trials: they heard a reference rhythm prompt (``X'') and were randomly presented with two clips (``A'' and ``B'') from the corresponding (1) TRIA generation, (2) MelodyFlow generation, or (3) a random drum excerpt to act as a low anchor. They were then asked to select ``A'' or ``B'' given the criteria:

\begin{displayquote}
    Select which of the two choices is a more musically pleasing translation from the reference clip to drums that captures the original rhythm and groove of the reference clip.
\end{displayquote}

Full coverage of our 80 sets required 3 pairwise comparisons per set: TRIA vs. MelodyFlow, TRIA vs. Random Excerpt, and MelodyFlow vs. Random Excerpt. We required 5 listeners evaluate each comparison, resulting in $80 \times 3 \times 5 = 1200$ total trials. From our ABX results, we computed the win rate of TRIA and MelodyFlow on each dataset and evaluated the statistical significance of the indicated listener preference. 
We present the results of our subjective evaluation in Figure \ref{fig:survey}. 

\textbf{Participant Recruitment}: We recruited 120 US English-speaking human listeners with an approval rating of $\geq95\%$ and a record of completing $100+$ prior tasks on Prolific. Each listener evaluated 10 randomly assigned ABX pairwise comparisons. To ensure data quality, participants had to pass a listening test assessing tone sensitivity from 55 Hz - 10 kHz \cite{rumbold2024correlations}, along with attention checks. They were paid \$2.50 per set of 10 comparisons, estimated to be equivalent to \$18.75/hour. We excluded participants who failed the listening test, as well as those who preferred the Random Excerpt $\geq80\%$ of the time, as this suggests they disregarded the given evaluation criterion of rhythm adherence. Following data cleaning, we had 116 participants with a total of 1160 evaluation pairs.

\subsection{Rhythm Prompt Adherence}

To evaluate TRIA's preservation of the rhythmic structure of sound gestures when translating to drums, we conduct an automated transcription evaluation. We sample 250 generations from each MelodyFlow and TRIA variant conditioned on rhythm prompts drawn from the AVP beatbox dataset, all of which have ground-truth human annotations of kick drum, snare, and hi-hat vocalizations. We then transcribe these generations using the pretrained ``Frame-RNN'' drum transcription model of Zehren et al. \cite{adtof}. Finally, we measure the correspondence between transcribed and ground-truth kick and snare drum parts using the onset F1 score with 30ms and 100ms tolerances, as is common in the drum transcription literature \cite{rnntranscription, beatnet}. Higher F1 scores indicate tighter correspondence between the kick and snare vocalizations in the rhythm prompt and the kick and snare drums in the generated audio. We report results in Table \ref{tab:transcription}.

\subsection{Timbre Prompt Adherence}

To evaluate TRIA's treatment of timbre information, we compute the cosine similarity between time-averaged 80-dimensional MFCC representations of the generated audio and timbre prompt (indicating adherence to the timbre prompt), the generated audio and rhythm prompt (indicating the degree of timbre leakage from the rhythm prompt), and the generated audio and a random excerpt from MoisesDB drums (as an anchor). While this measurement of spectral correspondence provides a coarse approximation of timbral similarity, we find that it captures strong trends in each model's treatment of timbre.  
Because MelodyFlow allows timbre specification through a text prompt, not an audio prompt, we can only compute its output similarity to the rhythm prompt and random excerpt. Our results are reported in Table \ref{tab:timbre}. We further illustrate the processing of rhythm and timbre prompts by both systems in Figure \ref{fig:uberfigure}.

\subsection{Audio Quality}

To evaluate the realism of generated audio, we compute the Kernel Audio Distance (KAD) \cite{kad, kid} between $500$ outputs from each method and a reference distribution of $500$ random excerpts of MoisesDB drums. Similar to Fr\'{e}chet Audio Distance (FAD) \cite{fad}, KAD measures the similarity of the generated distribution to a reference distribution, while showing stronger alignment with human quality ratings less bias at small sample sizes. For KAD we consider the ``PANN'' embedding variant \cite{pann2}, which the authors show is most correlated with human perception, and the ``CLAP-Laion-Music'' embedding variant \cite{clap_music}, which leverages a model trained specifically on music. Because TRIA receives audio timbre prompts from the reference distribution while MelodyFlow receives text timbre prompts, we compare only variants of TRIA for fairness. We report results in Table \ref{tab:quality}.

\section{Discussion}

Our experimental results demonstrate TRIA's efficacy in translating rhythm gestures to full-drumkit recordings  faithful to the rhythm and timbre prompts. As illustrated in Figure \ref{fig:survey}, our subjective evaluation shows no statistically significant preference between TRIA and MelodyFlow generations. \textit{This is promising given that MelodyFlow is roughly} $25 \times$ \textit{the size, and trained on} $2,000 \times$ \textit{the data}. 
Additionally, both models are strongly preferred to random drum excerpts at significance $p\leq0.001$ according to a two-sided binomial test, indicating that both succeed in capturing the core groove and structure of rhythm prompts.

\begin{figure}
  \centering
  \includegraphics[alt={TRIA versus MelodyFlow, visual comparison},width=0.99\linewidth]{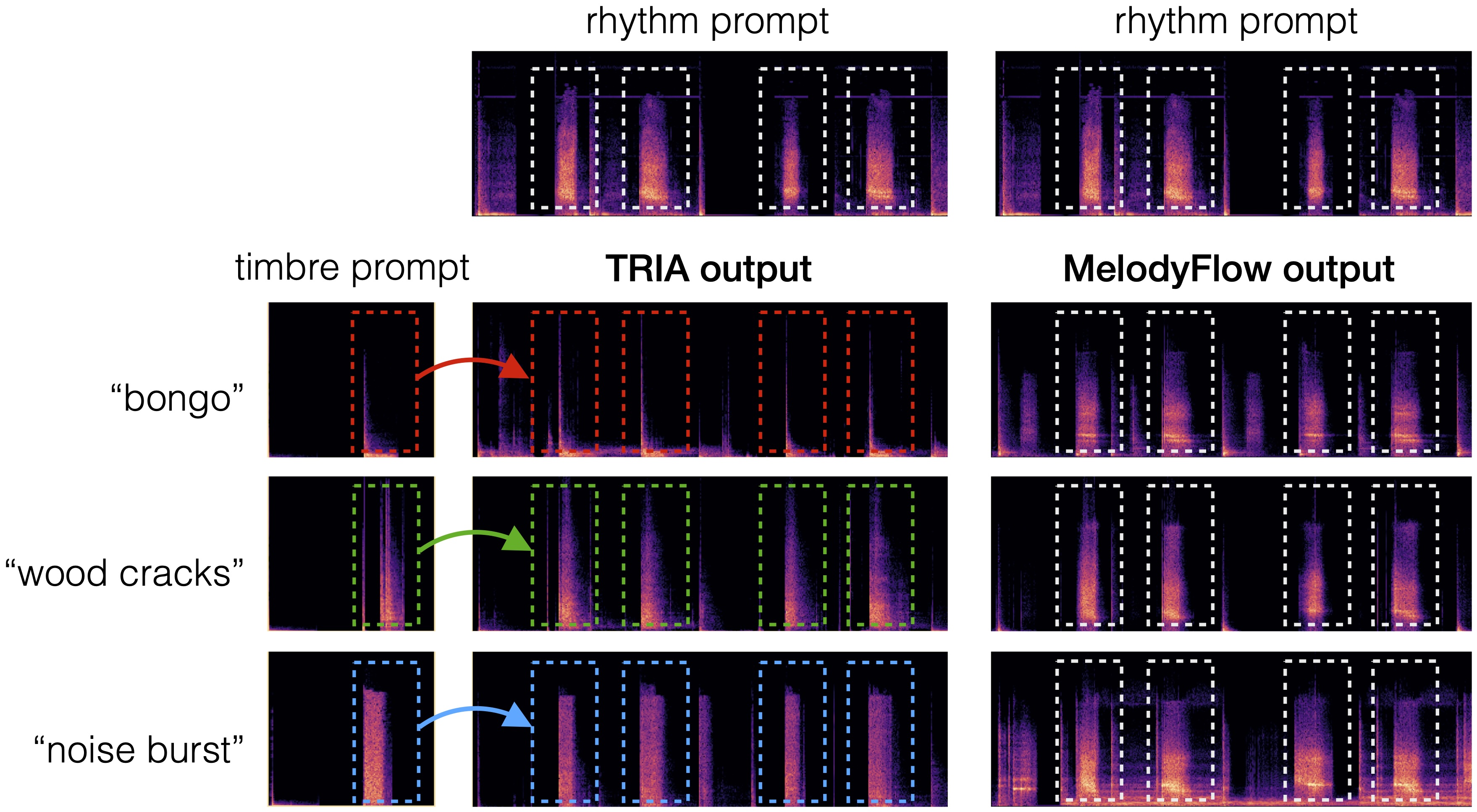}
  \caption{Given a rhythm prompt (top) with vocal kick and snare drum imitations, the “snare” sound can be replaced by user-provided samples via TRIA's timbre prompting ability: (a) a bongo drum, (b) wood cracks, and (c) a noise burst. Given corresponding timbre prompts in text form, MelodyFlow adheres more closely to the spectral content of the rhythm prompt.
  }
  \vspace{-1.0em}
  \label{fig:uberfigure}
\end{figure}

The results of our transcription evaluation, presented in Table \ref{tab:transcription}, show that TRIA strongly outperforms MelodyFlow in preserving the rhythmic structure of beatbox sound gestures as indicated by correspondence of kick and snare placement in the rhythm prompt and generated audio. While increasing the target flow step improves MelodyFlow's rhythm adherence slightly, it still significantly underperforms all evaluated TRIA variants. \textit{These results demonstrate the strength of TRIA's dualized rhythm feature representation, which outperforms both a 1-band representation and a non-adaptive 2-band representation that naively splits the mel spectrogram along its center frequency}. Adaptive 3- and 4-band rhythm feature representations yield diminishing returns as they slightly increase the accuracy of kick placement, but do not have a meaningful effect on snare placement. This indicates that a dualized representation may be sufficient to capture the core rhythmic structure of many sound gestures, while single-voice representations are likely insufficient.

The results of our timbre evaluation, presented in Table \ref{tab:timbre}, show that TRIA generations exhibit lower spectral correlation with the rhythm prompt than random anchors, and higher correlation with the timbre prompt than random anchors -- \textit{indicating both a lack of timbre leakage from the rhythm prompt and strong adherence to the timbre prompt}. In contrast, MelodyFlow generations exhibit higher-than-random spectral correlation with the rhythm prompt, indicating timbre leakage. We provide examples illustrating these behaviors in Figure \ref{fig:uberfigure}: MelodyFlow often mimics the spectral structure of rhythm prompts, while TRIA effectively utilizes a diverse array of timbre prompts to determine spectral structure. This audio-prompted timbre mapping is a key advantage of TRIA over text-prompted systems, allowing for more specific examplar-based steering of generations. Finally, as shown in Table \ref{tab:quality}, \textit{our dualized rhythm features outperform both 1-band and non-adaptive 2-band features in producing realistic drum audio}. 

Overall, these results show the promise of our proposed approach even in small model and data regimes. Directions for future work include scaling the model and dataset; leveraging TRIA's existing capabilities for other inference paradigms such as inpainting and drums-to-drums conversion; and exploring learnable dualized rhythm features.

\section{Acknowledgements}

This work was supported by NSF Award Number 2222369. We would also like to thank André Carvalho dos Santos and Gaël Le Lan for productive discussions.

\section{Ethics Statement}

In this section we acknowledge (1) the broader ethical implications of generative music models in the context of our work, (2) the ethical implications of using crowdworkers to perform our subjective evaluation, and (3) our positionality as authors of this work.

\subsection{Broader Ethical Implications}

A recent work on the ethical implications of generative audio models \cite{barnett2023ethical} identified a set of potential harms specific to generative music models: (1) loss of agency and authorship, (2) stifling of creativity, (3) predominance of western bias, (4) cultural appropriation, (5) copyright infringement, and (6) climate impact of these models; we address this work with regard to each of these six harms. 

This work is intended to provide creators with the ability to turn any sound gesture into a drum beat with their desired timbre. We see this as a means to provide music creators with additional agency; however, we do acknowledge that there is the (1) potential for removing agency or (2) stifling the creativity of percussion composition and production.

We recognize that our work is trained on a small dataset of drums and thus performs best with timbres present in that dataset, and so (3) may perform poorly with out-of-domain timbres such as traditional eastern music percussion instruments. This is a limitation of the dataset and current iteration of TRIA but not the proposed method itself, as future work could train TRIA on non-western drum beats to overcome this limitation. 

In its current iteration, we do not believe there is a strong potential for (4) cultural appropriation with TRIA; however, if someone were to re-train TRIA on a dataset of percussion from a culture to which they do not belong, it would enable that act. In regard to (5) potential for copyright infringement, TRIA was trained on MusDBHQ-18 \cite{musdb}, which is licensed for any educational purposes. If TRIA were to be used for commercial purposes, it would require re-training on proprietary datasets or otherwise non-copyrighted work in order to protect the copyright holders of these tracks, though we are not proposing this work be used for commercial purposes. 

Finally, we acknowledge (6) all generative models have an environmental impact---for transparency as encouraged by \cite{holzapfel2024green}, we documented our computational resources used for training, training time, and number of parameters, which in all cases are far less than needed for competing models such as MelodyFlow. Based on our $4 \times$ NVIDIA A10G GPUs (150W) and 27 hours of training time, we estimate each training run has an energy cost of 16.2 kWh. For comparison, MelodyFlow was trained on $8 \times$ H100 96GB GPUs (350W), with no reported training time. If we assume conservatively an equal training time of 27 hours, then one MelodyFlow training run would cost at least 280 kWh, or at a minimum $17 \times$ the energy consumption of TRIA.

\subsection{Crowdworkers}

Our subjective evaluation utilizing human listeners was approved (and determined to be exempt) under Institutional Review Board at the host university of the first author. We also ensured that each evaluator was paid a fair wage with an estimated hourly pay of \$18.75, which is above the minimum wage for every city in the United States. We also paid those who failed the listening test and thus could not partake in our study \$0.50 for their time. We used crowdworkers for this evaluation, and acknowledge that ethical use of crowdworkers goes beyond fair pay \cite{shmueli2021beyond}; we tested the study among the author team prior to launch to ensure there would be no burden to workers beyond potential boredom and made sure the evaluators knew they could stop the study at any time. 

\subsection{Positionality}

Finally, we would like to address the positionality of the authors. This is a diverse team of researchers, though we are predominantly from western developed countries (with one author being from the Global South). We are all both musicians and AI researchers, and thus share a mentality that AI technologies used for generative music can have a net positive impact as long as they are tools used to empower and assist musicians and creators rather than replace them. We acknowledge a bias in the conduct of this work reflecting an overall positive attitude towards AI technologies in this regard, and recognize that this is not a universal belief.

Ultimately, we believe that the benefits of this work far outweigh these potential risks, and we took care to keep them in mind as we conducted this research.

\bibliography{ISMIRtemplate}

\end{document}